\documentstyle[aps,prb,epsfig,floats]{revtex}
\begin{document}

\twocolumn[\hsize\textwidth\columnwidth\hsize\csname
@twocolumnfalse\endcsname

\title{A first-principles study of oxygen vacancy pinning of domain walls
in PbTiO$_3$}

\author{Lixin He and David Vanderbilt}
\address{Department of Physics and Astronomy, Rutgers University,
Piscataway, New Jersey 08854-8019}

\date{May 12, 2003}
\draft
\maketitle

\begin{abstract}
We have investigated the interaction of oxygen vacancies and
180$^\circ$ domain walls in tetragonal PbTiO$_3$ using
density-functional theory.  Our calculations indicate that the
vacancies do have a lower formation energy in the domain wall than
in the bulk, thereby confirming the tendency of these defects to
migrate to, and pin, the domain walls.  The pinning energies are
reported for each of the three possible orientations of the
original Ti--O--Ti bonds, and attempts to model the results with
simple continuum models are discussed.
\end{abstract}
\pacs{PACS: 61.72.-y, 85.50.-n, 77.84.-s}

\narrowtext

]

\section{INTRODUCTION}

Ferroelectric materials are of intense interest for use in
nonvolatile memory applications, in which the electric polarization
of an array element is used to store a bit of
information.\cite{jam,Auciello} However, the switchable
polarization tends to decrease after many cycles of polarization
reversal during device operation, a problem that is known as
polarization fatigue. In the last decade, polarization fatigue in
ferroelectrics has been under intensive study.  Although several
models have been proposed to explain this phenomenon,\cite{Tagan}
there is still no consensus, and many details of the fatigue
process remain unclear.  Nevertheless, it is generally believed
that defects, especially charged defects, play an important role.
For example, a series of experiments has provided some
understanding of how such defects may pin the domain
walls.\cite{Warren,Warren2}  In particular, attention has been drawn to
oxygen vacancies, which are often the most common and mobile
defects in perovskite ferroelectrics.  Moreover, it has been found
that fatigue resistance can be greatly improved by replacing the Pt
electrodes with conductive-oxide electrodes,\cite{Al-Shareef} which
can be explained in terms of the ability of oxide electrodes to control
the concentration of oxygen vacancies in the sample. Furthermore, the
fatigue rate in Pb(Zr,Ti)O$_3$ films was found to be very sensitive to
the oxygen partial pressure above the sample, suggesting
that oxygen vacancies strongly affect the fatigue process.\cite{Brazier}

Mechanisms for polarization fatigue based on pinning of domain
walls by oxygen vacancies have been discussed phenomenologically by
several authors.\cite{Tagan,Dawber,oxy-order}  However, in order to
put such phenomenologically theories on a firm atomistic basis, it
is important to have detailed first-principles calculations that
can provide information about the structure and energetics of the
ferroelectric domain walls, of the oxygen vacancies, and of the
interactions between the two.

While domain walls have been studied using Landau-type continuum
theories in earlier works,\cite{dw1,dw2,dw3} first-principles
calculations are essential for an accurate microscopic description
of the domain walls.\cite{Padilla,chadi-dw,meyer} Most recently,
Meyer and Vanderbilt studied 180$^\circ$ and 90$^\circ$ domain
walls in PbTiO$_3$ using first-principles methods, \cite{meyer}
establishing the geometry of the domain walls at the atomic
level and calculating the creation energy of the domain walls. The
domain wall was found to be extremely narrow, with a width of the
order of the lattice constant $a$; the positions of the atoms
change rapidly inside the domain wall and converge to
their bulk value very quickly outside.  As for oxygen vacancies,
recent calculations have provided very useful information about the
structure of oxygen-vacancy defects in this class of
materials.\cite{postnikov1,postnikov2,chadi}
However, to our knowledge, direct first-principles studies of the
interactions between vacancies and ferroelectric domain walls have
not yet appeared.

Thus, the goal of this work is to use first-principles calculations
to examine how an oxygen vacancy would interact with a ferroelectric
domain wall, and thus to shed some light on how oxygen vacancies
might affect the switching process and cause polarization fatigue.
Indeed, as the atomistic structure and polarization profile are
very different in a domain wall than in the bulk of a ferroelectric
domain, one may expect that the oxygen vacancies would also behave
differently in such very different environments.  To explore these
issues, we adopt PbTiO$_3$ as a model system for this study, and
calculate the formation energies for neutral oxygen vacancies
in the bulk and in 180$^\circ$ domain walls of tetragonal PbTiO$_3$.
Our calculations indicate that the vacancies do have a lower formation
energy in the domain wall than in the bulk, thereby confirming the
tendency of these defects to migrate to, and pin, the domain
walls.  The order of magnitude of the computed pinning energy is
$\sim10^{-1}\,$eV, with substantial variations depending on geometrical
configuration.

The rest of the paper is organized as follows. In Sec.~\ref{sec:method},
we describe the technical details of our computational method and the
supercells we used to model the domain walls. We present the results on
the vacancy pinning energies of each of the three possible orientations
of the original Ti--O--Ti bonds from first-principles calculations in
Sec.~\ref{sec:results}. A simple continuum model is introduced and used
to help understand these results in Sec.~\ref{sec:model}. In
Sec.~\ref{sec:discuss} we discuss briefly the anticipated role of
domain-wall pinning in fatigue, and discuss the case of charged
vs.\ neutral defects.  Finally, we summarize in
Sec.~\ref{sec:sum}.

\begin{figure*}
\begin{center}
   \epsfig{file=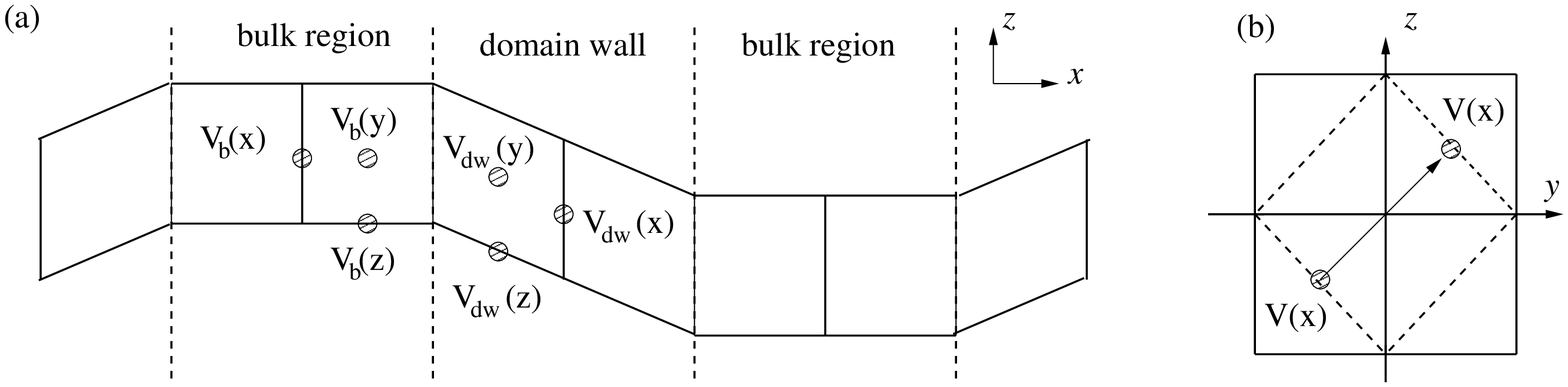,width=6.2in}
\end{center}
\caption{Schematic illustration of the 8$\times$$\sqrt{2}$$\times$$\sqrt{2}$
supercell used in the calculations.
(a) View in $x$-$z$ plane.  For purposes of illustration, the cell
is shown divided into ``bulk regions'' (polarized alternately along
$\pm\hat{z}$, not shown) and ``domain-wall regions'' (comprised of
the two primitive cells adjacent to the domain-wall center, which
lies on a Pb--O plane), although the actual behavior is slightly
more gradual.  V$_{\rm b}$ and V$_{\rm dw}$ refer to possible
locations of vacancies in bulk and domain-wall regions
respectively.
(b) View in $y$-$z$ plane (with $\sqrt{2}$$\times$$\sqrt{2}$
supercell illustrated by the dashed lines), showing the distance
between vacancies and their periodic images ($\sim a\sqrt{2}$).}
\label{fig:dw}
\end{figure*}

\section{METHODOLOGY}
\label{sec:method}

Our calculations are based on standard density-functional theory (DFT) within
the local-density approximation (LDA). We use a planewave pseudopotential
code implemented by the Vienna {\it ab-initio} Simulation Package
(VASP).\cite{kresse1,kresse2} Vanderbilt ultrasoft pseudopotentials
\cite{uspp} are used, with Pb $5d$ and Ti $3s$, $3p$ electrons
included explicitly.
The 29 Ry cut-off used here was well tested and determined to be adequate
for this material and these pseudopotentials.\cite{meyer}
The structure is considered to be relaxed when the forces are less than
0.05 eV/\AA; the change of total energy at this time is typically
less than 1 meV.

We first carried out reference calculations for domain walls without
vacancies, following closely the approach of Ref.~\onlinecite{meyer}.
We describe the 180$^\circ$ domain wall using an
8$\times$1$\times$1 supercell (long direction along $\hat{x}$) with
2$\times$4$\times$4 k-point sampling; the tetragonal axis and
polarization are along $\pm\hat{z}$, and two $yz$-oriented
domain walls divide the supercell into up and down domains of equal
width.

We also carried out reference calculations for vacancies in the bulk.
There are three kinds of oxygen vacancies that can be formed,
depending on whether the removed oxygen atom had its Ti--O bonds
along $\hat{x}$, $\hat{y}$, or $\hat{z}$, which we denote as
V($x$), V($y$), or V($z$), respectively.  These bulk vacancies were
studied using a 2$\times$2$\times$2 supercell with the tetragonal axis
(and polarization) chosen along $\hat{z}$, and with 2$\times$2$\times$2
k-point sampling.  (Thus, V($x$) and V($y$) are equivalent by
symmetry in the bulk.)  To obtain the vacancy energy, we first
calculate the total energy of the pure tetragonal
2$\times$2$\times$2 supercell as the reference energy. One oxygen
is then removed from the supercell in such a way that the supercell
remains net neutral, and the resulting structure is relaxed.  We
keep the volume and shape of the supercell fixed and allow only the
ion positions to relax. The vacancy energy is then calculated by
comparing the total energy difference before and after removing the
oxygen atoms from the supercell.

In order to study the interaction of the vacancies with the 180$^{\circ}$
domain wall, we constructed an 8$\times$$\sqrt{2}$$\times$$\sqrt{2}$
supercell by doubling the 8$\times$1$\times$1 supercell in the $y$-$z$
plane in a c$2$$\times$$2$ or $\sqrt{2}$$\times$$\sqrt{2}\,$R(45$^\circ$)
arrangement, and then removing one oxygen atom (and its periodic
images) from the interior of the domain-wall structure.  This
is illustrated schematically in Fig.~\ref{fig:dw}, referring here
to vacancies labeled ``V$_{\rm dw}$'' (i.e., located in the domain wall)
in Fig.~\ref{fig:dw}(a).  The vacancies form sheets in the $y$-$z$ plane,
with individual vacancies separated by a distance of about
$a\sqrt{2}$, as indicated in Fig.~\ref{fig:dw}(b).  Even
sparser arrangements would be desirable, but the supercell already
contains 80 atoms, and computational limitations make it difficult
to increase the separation further.  A 1$\times$3$\times$3 k-point
mesh is used for these supercells, and once again the ion positions
are allowed to relax while keeping the supercell volume and shape
fixed.  (A single isolated domain wall decorated by vacancies could
not expand or contract in the $\hat{y}$ or $\hat{z}$ directions
because of the epitaxy constraint to the bulk, and our experience,
consistent with Ref.~\onlinecite{meyer}, indicates that the lattice
would expand very little in the $\hat{x}$ direction if allowed to
do so.)

Finally, in order to calculate the energy difference for an oxygen
vacancy to be inside the domain wall, relative to being in bulk, we
find that it is advantageous to carry out corresponding
8$\times$$\sqrt{2}$$\times$$\sqrt{2}$ supercell calculations in
which the oxygen vacancy has been removed from a bulk-like region
of the supercell.  In this way, we reduce the systematic errors
associated with supercell shape, k-point sampling, etc.  Our
results for the binding energies of oxygen vacancies in the domain
walls will normally be based on calculations of this kind, except
where unavailable as noted below.

\section{RESULTS}
\label{sec:results}

We first report our calculations on isolated vacancies in bulk 
ferroelectric PbTiO$_3$ in the 2$\times$2$\times$2 supercell.
Using the theoretical values of the lattice constants
($a=3.86$ \AA, $c/a=1.0466$) as obtained in Ref.~\onlinecite{meyer},
our calculations show that the oxygen vacancies of type V($z$)
are more stable than V($x$) and V($y$) by 0.3 eV, similar to what
was found by Park and Chadi for somewhat different conditions.\cite{chadi}
As indicated in Sec.~\ref{sec:method},
however, we prefer to calculate pinning energies for oxygen
vacancies in domain walls by using reference bulk vacancy
calculations in an 8$\times$$\sqrt{2}$$\times$$\sqrt{2}$ supercell
that can be used with and without domain walls, in order to provide
maximum cancellation of systematic errors.  Such calculations will
be the basis for the the results given in this section.  However,
we will return to the use of the 2$\times$2$\times$2 supercell in
Sec.~\ref{sec:model} for some calculations relevant to the modeling
of vacancies in a perturbed environment.

We then confirm the structure of the relaxed 180$^\circ$ domain wall.
Our results for the 180$^{\circ}$ domain wall are very close to those
of the previous calculation.\cite{meyer}
The domain wall is located on the Pb-O plane.
The atomic displacements converge rapidly to their bulk values,
with most of the displacements ocurring within one unit cell of the
domain wall center.  Consequently, the polarization reverses sharply
within approximately one unit cell of the domain wall, and quickly
saturates to a bulk value further away.  Thus, the domain wall is extremely
narrow, only about two lattice constant wide, and we find that it
displays a significant $x$-$z$ shear in addition to the reduced
polarization.   With this justification,
we can heuristically divide the supercell into two different regions,
a bulk region and a domain-wall region, as sketched in oversimplified
form in Fig.~\ref{fig:dw}(a).

To minimize the interactions between neighboring vacancies, the domain-wall
supercell is doubled as described in Sec.~\ref{sec:method}.
The shape of the supercell in the $y$-$z$ plane is indicated by
the dashed lines in Fig.~\ref{fig:dw}(b).
The total energy E$_0$ of this domain wall structure (80-atom
$8$$\times$$\sqrt{2}$$\times$$\sqrt{2}\,$
supercell) is calculated for use as the reference energy.

To calculate the oxygen vacancy energy, we remove one oxygen
from the supercell and relax the resulting structure.
Recall that there are three types of oxygen vacancies, V($x$), V($y$), and
V($z$), according to the orientation of the Ti--O--Ti bond from which
the oxygen atom has been removed.  For each type, we choose one vacancy
as close as possible to the center of the bulk region, and another
as close as possible to the central domain-wall plane.  We label these
as `b' (bulk) and `dw' (domain wall), respectively.
Thus, V$_{\rm dw}$($x$)
is an $x$-oriented vacancy at the domain wall, etc.
In all calculations, we keep the supercell charge-neutral.

All unrelaxed vacancies have an $M_y$ mirror symmetry
through the defect site.  In addition, the
unrelaxed V$_{\rm dw}$($x$) and V$_{\rm b}$($x$) defects have extra
$C_2^{(y)}$ and $M_x$ symmetries respectively, because they
lie precisely in, or else half-way between, the domain-wall planes.
When relaxing the structure, we observe that the two Ti ions that neighbor
the vacancy site relax most, as expected.
In the bulk region, the relaxation of V$_{\rm b}$($x$) and
V$_{\rm b}$($z$) does not lead to any symmetry breaking, while for
V$_{\rm b}$($y$) there are Pb ion displacements that break the
$M_y$ symmetry and contribute about 10 meV to the relaxation energy.
In the domain-wall region, V$_{\rm dw}$($y$) also has
similar distortions, but the energy is only lowered by about 5 meV.
As for V$_{\rm dw}$($x$), which starts with a relatively high
unrelaxed symmetry (four-element point group C$_{2h}$), the
relaxation breaks the $M_y$ and $C_2^{(y)}$ symmetries so that the
only remaining symmetry is inversion (point group C$_{1h}$), and
the total energy is lowered by about 20 meV.

The total energy of the oxygen-vacancy supercell is then calculated
for each configuration, and a vacancy formation energy is computed
using an appropriate reference.  The results are listed in
Table~\ref{tab:comb-energy}.  Here
$\Delta E_{\rm dw} = E_{\rm v\in dw}+E_{\rm oxy}-E_{\rm dw}$,
where $E_{\rm v\in dw}$ is the energy of the supercell with the
vacancy in the domain wall, $E_{\rm oxy}$ is the energy of a free
oxygen atom, and $E_{\rm dw}$ is the energy of a vacancy-free
supercell with domain walls; and either
$\Delta E_{\rm b}=E_{\rm v+dw}+E_{\rm oxy}-E_{\rm dw}$
(``Ref.\ with DW'') or
$\Delta E_{\rm b}=E_{\rm v}+E_{\rm oxy}-E_0$
(``Ref.\ without DW''), where $E_{\rm v+dw}$ is the energy of a
vacancy in the bulk-like region of a supercell containing domain
walls, $E_{\rm v}$ is the energy of a vacancy in a domain-wall-free
supercell, and $E_0$ is the energy of the corresponding bulk supercell
containing neither domain walls nor vacancy.

\begin{table}
\caption{Vacancy formation and pinning energies.  $\Delta E_{\rm dw}$
is the energy to create a vacancy in the domain-wall region of the
supercell.  $\Delta E_{\rm b}$ is the reference energy to create a
vacancy either in a supercell without domain walls (``Ref.~without
DW''), or in the bulk-like region of a supercell with domain walls
(``Ref.~with DW''), and $E_{\rm p}=\Delta E_{\rm b}-\Delta E_{\rm dw}$
is the corresponding pinning energy.  All energies are in eV.}
 \label{tab:comb-energy}
\vskip 0.1cm
\begin{tabular}{cddddd}
 & & \multicolumn{2}{c}{Ref.~without DW} & \multicolumn{2}{c}{Ref.~with DW} \\
 & $\Delta E_{\rm dw}$ &
   $\Delta E_{\rm b}$ & $E_{\rm p}$ &
   $\Delta E_{\rm b}$ & $E_{\rm p}$ \\
\hline
\multicolumn{6}{l}{Unrelaxed} \\
\quad V($x$) & 10.726 & 10.777 & 0.051 & 10.764 & 0.038 \\
\quad V($y$) & 10.943 & 10.946 & 0.003 & 10.929 & $-$0.014 \\
\quad V($z$) & 10.939 & 11.102 & 0.163 & 11.098 & 0.159 \\
\multicolumn{6}{l}{Relaxed} \\
\quad V($x$) & 10.445 & 10.567 & 0.122 & 10.542 & 0.097 \\
\quad V($y$) & 10.660 & 10.743 & 0.083 & 10.736 & 0.076 \\
\quad V($z$) & 10.459 & 10.720 & 0.261 &  ---   &  ---  \\
\end{tabular}
\end{table}

As shown in Table~\ref{tab:comb-energy}, all the vacancies in the
domain wall have lower formation energies than their counterparts
in the bulk.  This indicates that there is an attractive
interaction between the vacancy and the domain wall and implies a
positive pinning energy $E_{\rm p}=\Delta E_{\rm b}-\Delta E_{\rm dw}$.

Where available, the values for $E_{\rm p}$ obtained using the
reference supercell with domain walls is to be preferred (last two
columns of Table~\ref{tab:comb-energy}), since one expects a more
systematic cancellation of errors in this case.  However, in some
situations this turns out not to be possible.  For the vacancy
V($z$), in particular, a problem arises.  We find that when we
attempt to compute the relaxed energy $E_{\rm v+dw}$ of the
supercell in which the vacancy has been placed as far from the
domain walls as possible, the domain wall actually shifts its
position during the relaxation in order to coincide with the vacancy,
thus spontaneously converting the supercell from $E_{\rm v+dw}$
to $E_{\rm v\in dw}$.  This will be discussed further in
Sec.~\ref{sec:dw_shift}.  For this case, our best value for
$E_{\rm p}$ of 261 meV is obtained by falling back to the use of
a reference supercell without domain walls (middle columns of
Table~\ref{tab:comb-energy}).  By looking at other cases
(i.e., V($x$), V($y$), and unrelaxed cases), it can be seen that
an uncertainty of approximately 25 meV is introduced by the
use of the less preferable reference supercell.

It may be noted in Table~\ref{tab:comb-energy} that the formation
energies $\Delta E_{\rm b}$ are significantly different
for V$_{\rm b}$($x$) and V$_{\rm b}$($y$) ($\sim$200 meV),
although by symmetry they should be equal in a true bulk
environment.  Here the differences arise from a
supercell size effect connected with the arrangement of vacancies
into sheets of fairly high density in the $y$--$z$ plane.
For a sufficiently large supercell we would expect these energies to
become equal, because the local environments of V$_{\rm b}$($x$) and
V$_{\rm b}$($y$) would be almost
identical and the interactions between them would be negligible.
However, in our case the vacancies are only separated by a distance
of about $a\sqrt{2}$, so the interactions are not
negligible. On the other hand, the vacancies in the domain walls
should have similar interactions, and we can expect some cancellation
of errors when arriving at the pinning energy.  Thus, we have more
confidence in the $E_{\rm p}$ values than in the relative
formation energies of V($x$), V($y$) and V($z$).

\subsection{Domain-wall shift}
\label{sec:dw_shift}

\begin{figure}
\begin{center}
   \epsfig{file=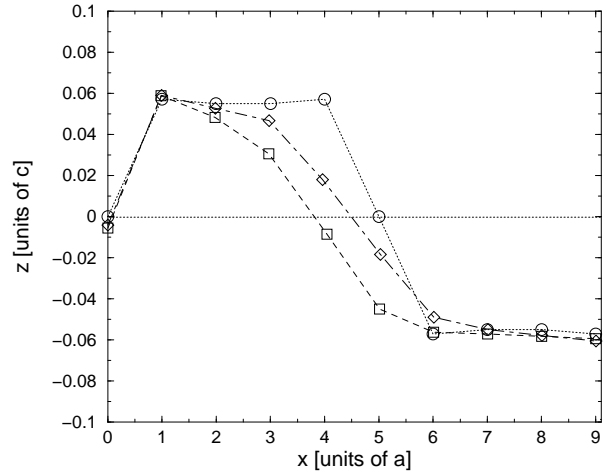,width=3.1in}
\end{center}
\caption{Circles: layer-by-layer Pb-atom $z$ coordinates for a relaxed
$10$$\times$$\sqrt{2}$$\times$$\sqrt{2}$ supercell with domain walls
centered at $x=0$ and $x=5a$.  Diamonds: same, but with a layer of
V($z$) vacancies inserted at $x=4.5a$. Squares: same, but with the
vacancy layer at $x=3.5a$.  The domain wall can be seen to
shift toward the vacancies.}
\label{fig:shift}
\end{figure}

As indicated in the previous subsection, when we attempt to calculate
the energy $E_{\rm v+dw}$ of the V$_{\rm b}$($z$) vacancy in the
bulk-like region of an $8$$\times$$\sqrt{2}$$\times$$ \sqrt{2}$ supercell
containing domain walls, the domain wall spontaneously shifts
toward the vacancy during the relaxation process.  We carried out
further tests using a $10$$\times$$\sqrt{2}$$\times$$\sqrt{2}$
supercell and observed the same phenomenon, as shown in
Fig.~\ref{fig:shift}.  First, we put the vacancy on a TiO$_2$ plane
inside the domain wall ($x=4.5a$) and allow relaxation.  We observe
that the Pb-centered domain wall shifts towards the vacancy and
becomes a Ti-centered domain wall.  We then put the vacancy on a
TiO$_2$ plane near the domain wall ($x=3.5a$), and observe that the
domain wall center (originally at $x=5a$) again shifts to the left
and becomes centered roughly at the TiO$_2$ plane at $x=3.5a$,
ending up with almost same total energy as before.  When we attempt
to put the vacancy even farther from the domain wall at $x=2.5a$, we
find that a new domain wall forms at the vacancy position.
Clearly, the domain wall is trying to shift its position in each case
so as to minimize the polarization at the position of the vacancy,
thereby demonstrating directly the pinning effect of oxygen vacancies
and ferroelectric domain walls.

This effect is most pronounced for the case of V$_{\rm b}$($z$)
because it has the strongest pinning energy, as can be seen in
Table~\ref{tab:comb-energy}.  In the case of V$(y)$, we find a
similar but weaker effect.  That is, if the V$(y)$ vacancy is
placed close enough to the domain wall (e.g., at $x=4.5a$ in
Fig.~\ref{fig:shift}), a similar shift can occur; but no shift
occurs if the defect is placed farther away.

To pursue the calculation of $E_{\rm v+dw}$ in order to obtain
$\Delta E_{\rm b}$ for V$_b$($z$), it would be necessary for us to
use a supercell larger than $\sqrt{2}$$\times$$\sqrt{2}$ in the
$y$--$z$ plane.  However, as this would be computationally
prohibitive, we have instead chosen to recalculate the
bulk vacancy energies using an $8$$\times$$\sqrt{2}$$\times$$\sqrt{2}$
supercell without domain walls, as indicated in the previous
subsection.  This provides an alternative reference energy which,
though less accurate, is available in all cases.

\section{MODEL CALCULATION}
\label{sec:model}

Our first-principles calculations give us a rough picture of the interactions
between oxygen vacancies and domain walls. These calculations show that
the domain walls can indeed be pinned by oxygen vacancies.
We obtain estimates for the pinning energies, and find that the
pinning energy for V($z$) is much larger than for V($x$) or V($y$).
We would like to understand better the physics underlying these results,
and to appreciate which results might generalize to other
situations (e.g., other ferroelectrics materials, or other
domain-wall structures such as 90$^\circ$ boundaries).

With this motivation, we consider a model in which the vacancy
formation energy depends on the immediate vacancy environment as
characterized by local polarizations and strains.  In particular,
we consider a continuum description of the polarization and strain
fields in the domain wall, as would occur in a Landau-type model,
and assume that the vacancy energy can be expressed as a function
of the local strain and polarization only.  (A more sophisticated
model might involve also a dependence on the local gradients of
these fields, but we have not pursued this here.) Thus, in general
we would write the vacancy formation energy as $E_{\rm v}(\eta,
{\bf P})$, where $\eta$ and $\bf P$ are the strain tensor and
polarization vector describing the state of the
local environment before removal of the oxygen atom.  However, we
specialize here to the case of interest, a 180$^\circ$ wall lying
in a $y$--$z$ plane separating tetragonal phases with polarizations
along $\pm\hat{z}$ on either side.  Thus, we focus on only the $z$
component of polarization, and for convenience we define a
dimensionless reduced polarization $p=P_z/P_{\rm bulk}$.  As for
the strain tensor, we have $\eta_{yy}=\eta_{yz}= \eta_{zz}=0$ by
lattice continuity and $\eta_{xy}=0$ by symmetry.  Moreover, we
find that $\eta_{xx}$ remains quite small in the domain-wall
region.  Thus, we focus only on the $xz$ shear strain component
and let $\eta_s=(a/c)\,\eta_{xz}$.  The vacancy formation energy
$E_{\rm v}(\eta_s,p)$ is thus considered as a function
of the local $\eta_s$ and $p$.

Expanding $E(\eta_s, p)$ in powers of $p$,
\begin{equation}
E_{\rm v}(\eta_s, p)=A(\eta_s)+ B(\eta_s) \, p^2 + {\cal O}(p^4) \, ,
\label{eq:fit}
\end{equation}
where odd powers in $p$ have been dropped by symmetry, and
only terms up to ${\cal O}(p^2)$ are retained henceforth.
We then expand $A(\eta_s)$ and $B(\eta_s)$ in powers of
$\eta_s$ as
\begin{equation}
\label{eq:fita}
A(\eta_s)=a_0 + a_1 \eta_s + a_2 \eta_s^2 + {\cal O}(\eta_s^3)
\end{equation}
and
\begin{equation}
\label{eq:fitb}
B(\eta_s)=b_0 + b_1 \eta_s + b_2 \eta_s^2 + {\cal O}(\eta_s^3) \, .
\end{equation}
Dropping terms beyond quadratic order in $\eta_s$, we take the
coefficients $a_1$, $a_2$, $a_3$, $b_1$, $b_2$ and $b_3$, in
Eqs.~(\ref{eq:fit}-\ref{eq:fitb}) to constitute the parameters
of our model.

\begin{figure}[t!]
\begin{center}
   \epsfig{file=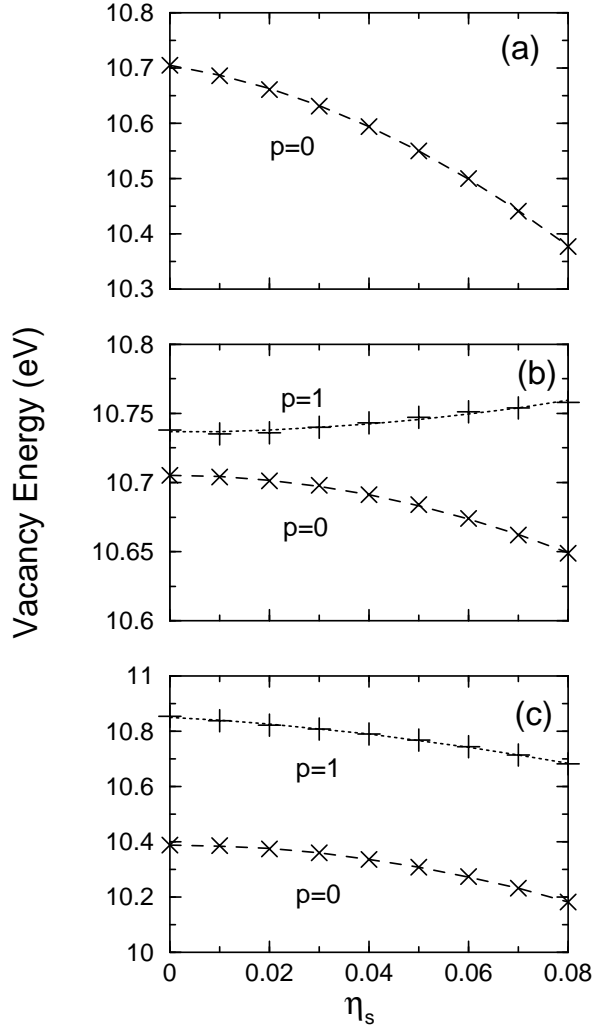,width=3.1in}
\end{center}
\caption{Symbols: calculated vacancy formation energies
{\it vs}.\ shear strain as obtained from $2$$\times$$2$$\times$$2$
supercell calculations for (a) V($x$), (b) V($y$), and (c) V($z$).
Plus signs and crosses are for $p=1$ and $p=0$ respectively.  Lines
are fits to Eqs.~(\protect\ref{eq:fit}-\protect\ref{eq:fitb}).}
\label{fig:fit}
\end{figure}

To obtain these six parameters, we do a series of
calculations at a set of different $\eta_s$ values for both $p=0$
and $p=1$. To do this, we construct $2$$\times$$2$$\times$$2$
supercells at different fixed values of $\eta_s$ and calculate the
vacancy energies as in the last section, allowing relaxation of
ionic positions but not strains.  A $2$$\times$$2$$\times$$2$
k-mesh is used in all these calculations.  We do calculations at
$p=0$ by enforcing inversion symmetry of the lattice.  If there is
no constraint of symmetry imposed, the ions relax freely, resulting
in $p=1$.  The results of these calculations are shown in
Fig.~\ref{fig:fit}.

It might be thought that the coefficients $a_1$ and $b_1$ in
Eqs.~(\ref{eq:fita}-\ref{eq:fitb}) should vanish by symmetry, but
if the oxygen vacancy induces a distortion which lowers the lattice
symmetry as discussed in Sec.~\ref{sec:results}, this may not
always be true.  Consider, for example, the case of V($x$) at
$\eta_s=0$ and $p=0$ (inversion symmetry imposed).  While the
unrelaxed defect has D$_{2h}$ symmetry, the relaxation lowers the
symmetry to C$_{2h}$ ($E$, $I$, $M_y$ and $C_2^y$) and reduces
the energy by about 65 meV relative to the case where no symmetry
breaking is allowed.  Actually, there are two equivalent defects,
related to each other via the broken symmetry $M_x$, having
degenerate energies at $\eta_s=0$ but having $a_1$ coefficients of
opposite sign (i.e., opposite response to an applied $xz$ strain
$\eta_s$.)  In Fig.~(\ref{fig:fit}) only the energy of the more
stable of these two defects is plotted as a function of $\eta_s>0$,
and the dashed line is a fit to Eq.~(\ref{eq:fita}).  The resulting
(negative) value of $a_1$ is given in Table \ref{tab:fit}.

Considering the other vacancies at $p=0$,
vacancy V($y$) has a similar symmetry breaking but its
orientation is such that the degeneracy would be split by a
$\eta_{yz}$ strain, which is absent here. For V($z$) we observe no
symmetry breaking at $p=0$.  Thus $a_1$ vanishes for these cases.

Turning now to $p=1$, we find a reversed situation: V($x$) and
V($y$) show no breaking of their C$_{2v}$ symmetry at $\eta_s=0$,
while V($z$) breaks from C$_{4v}$ to C$_{1h}$ after relaxation.
Thus, $b_1=0$ for V($y$) but not for V($z$).  For non-zero
$\eta_s$ in Fig.~(\ref{fig:fit}), no further symmetry breaking is
observed beyond what was already present at $\eta_s=0$.

The parameters resulting from all the fits of
Eqs.~(\ref{eq:fit}-\ref{eq:fitb}) are listed in Table \ref{tab:fit}.
From the fact that $a_1$ and $a_2$ are negative,
we see that the vacancies will prefer an environment of high shear
strain.  Similarly, since $b_0$ is positive, they will prefer an
environment of low polarization.  Thus, the parameters are suggestive
of a tendency for the vacancies to pin the domain walls.

\begin{table}
\caption{Coefficients obtained by fitting to the model calculations.
(For V($x$), $b_1$ and $b_2$ are not needed for the pinning energy
and thus are not reported.)}
 \label{tab:fit}
\vskip 0.1cm
\begin{tabular}{cdddddd}
       & $a_0$   & $a_1$    &$a_2$     &$b_0$    &$b_1$     &$b_2$\\
\hline
V($x$)   & 10.705  & $-$1.434   & $-$33.117  & 0.032\\
V($y$)   & 10.705  & 0        &  $-$8.737  & 0.032 & 0      & 12.315\\
V($z$)   & 10.389  & 0        & $-$32.008  & 0.462 & $-$0.994 & 18.307
\end{tabular}
\end{table}

In order to model quantitatively the vacancy formation energy in the
domain wall, we now have to estimate the values of $p$ and $\eta_s$
that occur in a vacancy-free domain wall at the location where the
vacancy would occur.  Since V$_{\rm dw}$($x$) lies in the
Pb-O symmetry plane, $p=0$ there.  From the first-principles
calculations of Ref.~\onlinecite{meyer}, we can estimate that
$p\simeq0.8$ at the neighboring TiO$_2$ plane where V$_{\rm dw}$($y$)
and V$_{\rm dw}$($z$) are located.

The estimation of $\eta_s$ is more subtle.  The problem is that the
shear strain is not well defined in the domain walls, since it depends
strongly on which of the sublattice we follow. For example, if we
define the shear strain to be $\eta_s=\delta_z/c$, where $\delta_z$, is
the displacement of the adjacent atoms of same type in the $\hat{z}$
direction, we estimate that $\eta_s({\rm Pb})=0.068$,
$\eta_s({\rm Ti})=0.054$, $\eta_s({\rm O(x)})=-0.029$,
$\eta_s({\rm O(y)})=-0.047$, and $\eta_s({\rm O(z)})=-0.027$ in
the center of the domain wall.  This variation reflects the reversal
of the polarization-related displacements along $\hat{z}$ as one passes
through the domain wall along $\hat{x}$.  We could define a mean shear
as $\hat{\eta}_s=(1/5)\sum_i{\eta_s(i)} \approx 0.005$, which as expected
is about half of the ``geometrical offset'' defined in
Ref.~\onlinecite{meyer} (the offset occurs over approximately two
unit cells).  An alternative choice is the root-mean-square shear
strain $\bar{\eta}_s=[(1/5) \sum_i \eta_s^2(i)]^{1/2}=0.048$.
We expect that the most reasonable choice of an effective
$\eta_s^{\rm eff}$ should lie somewhere between these two limits.
For V$_{\rm dw}$($x$), the pinning energy we get from first-principles
calculation is 97 meV.  Comparing with Eqs.~(\ref{eq:fit}-\ref{eq:fitb}),
we find that $\eta_s^{\rm eff}=0.03$ gives a reasonable agreement with
the first-principles result for this case, and we thus adopt this value.
The shear strain should be slightly smaller at the location of
V$_{\rm dw}$($y$) or V$_{\rm dw}$($z$), half a lattice constant
away from the domain-wall center, but for simplicity we retain the
same value of $\eta_s=0.03$ for all three defects.

Using the parameters from Table \ref{tab:fit} and the values of
$\eta_s$ and $p$ discussed above, we may estimate the
pinning energy $E_{\rm p}=E_{\rm v}(0,1)-E_{\rm v}(\eta_s,p)$ via
\begin{equation}
\label{eq:pinning}
 E_{\rm p} = (1-p^2)b_0  - (a_1+b_1p^2) \eta_s-
(a_2+b_2p^2)\eta_s^2 \;  .
\end{equation}
The resulting pinning energies are reported, and compared with the
direct first-principles calculations, in Table \ref{tab:compar}.

\begin{table}
\caption{Environmental model parameters appearing in
Eq.~\protect\ref{eq:pinning}, and resulting pinning energy $E_{\rm p}$
of the model compared with the best estimate from the direct
DFT calculations in Table \protect\ref{tab:comb-energy}.}
 \label{tab:compar}
\vskip 0.1cm
\begin{tabular}{ldddd}
& $\eta_s$ & $p$ & $E_{\rm p}^{\rm model}$ (eV) & $E_{\rm p}^{\rm DFT}$ (eV) \\
\hline
V($x$)   & 0.03 & 0.0 & 0.105 & 0.097 \\
V($y$)   & 0.03 & 0.8 & 0.012 & 0.076 \\
V($z$)   & 0.03 & 0.8 & 0.187 & 0.261 \\
\end{tabular}
\end{table}

Recall that the good agreement for V($x$) occurs by
construction.  The agreement for V($z$) is fair and
that for V($y$) is somewhat poor, although at least
we have the correct sign of the pinning energy even in the worst
case of V($y$).  Thus, we find that the present system
is sufficiently complex that our simple model description is only
partially successful.

There are several reasons why this may be.  As discussed in
Sec.~\ref{sec:dw_shift}, the domain walls may shift if the density
of oxygen vacancies is high, and the shift increases the pinning
energy.  This may help explain the underestimation of the pinning
energies for V($y$) and V($z$).\cite{explan-t} Also, the use of a
2$\times$2$\times$2 supercell for the calculations of Fig.~\ref{fig:fit},
from which the values of Table \ref{tab:fit} were obtained, means that
the vacancies were much closer to the dilute limit than was the case
for the vacancy-in-domain-wall calculations of Table \ref{tab:comb-energy}.
This is most likely the dominant source of the discrepancy for the case of
V$_{\rm dw}$($y$), which is less sensitive to $\eta_s$.
Indeed, the fact that a higher density of oxygen vacancies in the
domain wall leads to a larger pinning energy is consistent with a picture
in which oxygen vacancies would tend to make planar clusters in the
domain wall, thus acting to increase the pinning energy. \cite{oxy-order}
Future tests on larger supercells in the $y$--$z$ directions might help
clarify these issues, although these still remain intractable for the
time being.

Another obvious source of the discrepancies may be the limitations
of the model.  It is unsatisfying that the choice of $\eta_s$ is
so ambiguous, and it is unclear whether two variables ($\eta_s$ and
$p$) should suffice to describe the local environment.
After all, the structural distortions change rapidly as one passes through
the domain wall, so that it is not clear whether a Landau-type continuum
model should be expected to capture the details of the energetics.
It might be interesting to see whether a model more like the
effective-Hamiltonian description of Bellaiche
{\it et al.},\cite{Bellaiche,Hemphill,George} which includes
compositional disorder in order to treat alloys, could be
successfully used here.

Nevertheless, we believe that our model captures some of the essential
physics of the pinning mechanism of oxygen vacancies in the 180$^\circ$
domain walls.  It gives the correct sign and overall order of magnitude
for the pinning energy, and correctly reflects that the pinning is
strongest for V$_{\rm dw}$($z$), intermediate for V$_{\rm dw}$($x$),
and weakest for V$_{\rm dw}$($y$).  It also helps clarify the relative
roles of strain and polarization effects in the pinning mechanism.
We thus expect that it may be of some use for understanding other
ferroelectric materials as well.

\section{DISCUSSION}
\label{sec:discuss}

We now briefly discuss how oxygen vacancies may affect the
ferroelectric switching process.  As is well known, switching in
ferroelectrics occurs not through a homogeneous concerted reversal
of the polarization in the bulk, but through the motion of domain
walls separating regions of different polarization.  Thus, insofar
as these domain walls become pinned, the switching will be
suppressed.

In a pristine ferroelectric material that has a robust hysteresis
loop and a large remanant polarization, the ferroelectric
domain walls are presumably only weakly pinned by some
pre-existing defects.  Our calculations indicate that oxygen
vacancies will tend to migrate into these domain walls over
time, since they experience a binding energy to the domain wall
of between 100 and 250 meV.  In fact, once inside the domain wall, we would
expect the vacancy to hop into the V($z$) configuration, since
this is lower in energy than the V($x$) or V($y$) configurations.
Thus, the effective binding energy is $\sim$250~meV, the value
associated with the V($z$) vacancy.

If a significant number of vacancies accumulate in the domain wall,
they in turn can act to pin the domain wall.  When the density of
such vacancies is low, they will not pin the domain walls strongly,
and switching will still be able to occur.  As the areal density of
vacancies increases, however, an increasing fraction of domain
walls (or increasingly large portions of individual domain walls)
may become immobile, resulting in the decay of the switchable
polarization.

Of course, there are many limitations of our theoretical analysis,
and the real experimental situation could be much more complicated.
We have studied neutral oxygen vacancies, which correspond to
vacancies of charge +2$e$ neutralized by electrons residing in nearby
states of mainly Ti $3d$ character.  In the absence of neutralizing
electrons, the vacancies may pin the domain walls even more strongly.
(In this case, the situation becomes more complicated, since we can
expect that the domain walls may acquire a tilt in order to
compensate the vacancy charges.  That is, if the 180$^\circ$ domain
wall is not exactly parallel to the polarization, there is a bound
charge $\Delta{\bf P}\cdot\hat{n}$ that can help neutralize the
vacancy charges, an effect which may contribute to the strength of
the pinning effect.\cite{Warren2})  Alternatively, the oxygen
vacancies may tend to aggregate into clusters\cite{oxy-order} or to
form defect complexes of various kinds.  Finally, the polarization
switching process is rather a complicated dynamical process that we
have not attempted to model in detail.

Nevertheless, we believe that our first-principles results serve as
a first step towards understanding the possible role of oxygen
vacancies in the pinning of ferroelectric domain walls.  While we
have investigated only one class of defects that may be involved in
pinning, at least our calculations provide a lower limit for the
strength of the pinning effect, which may be stronger if other defects
or defect complexes play the dominant role.  Our results may also
serve as input for more complex modeling and simulation.  For
example, it could be used to extend a model such as that of
Ahluwalia and Cao, who have done simulations of domain-wall
formation in a 2D model simulation,\cite{Ahluwalia} by the
inclusion of vacancies into the simulation.  Our results might also
be useful in the formulation of an effective-Hamiltonian
approach\cite{zhong} that could be used to carry out
finite-temperature simulations of the domain-wall behavior.  Such
studies might help quantify the effective strength of the pinning
effect under more realistic conditions.

\section{SUMMARY}
\label{sec:sum}

In summary, we have used first-principles density-functional calculations
to investigate the interaction of oxygen vacancies and 180$^\circ$ domain
walls in tetragonal PbTiO$_3$.  Our calculations indicate that the
vacancies do have a lower formation energy in the domain wall than
in the bulk, thereby confirming the tendency of these defects to
migrate to, and pin, the domain walls. The pinning energies are
calculated for each of the three possible orientations of the
original Ti--O--Ti bonds, and are found to be 97 meV, 76 meV and 261 meV for
V($x$), V($y$) and V($z$) respectively. We also introduce a simple continuum
model with only two parameters ($p$, $\eta_s$) to model the results.
This simple model gives pinning energies that agree qualitatively with
the first-principles calculations, and we expect that it may
prove useful for other ferroelectric systems as well.

This work was supported by ONR grant N0014-97-1-0048.



\begin{references}
\bibitem{explan-t}
If we insert $p$=0 into Eq.~(\ref{eq:pinning}), reflecting the
idea that the domain wall may shift to become centered on the
vacancy layer, we find $E_p$=0.040 and 0.490 eV for V($y$)
and V($z$) respectively.  This improves the agreement for
V($y$), but overshoots for V($z$), in comparison with the
direct DFT results.
\end{references}
\end{document}